\documentclass[twocolumn,secnumarabic,amssymb, nobibnotes, aps, prd,superscriptaddress]{revtex4-1}

\usepackage{graphicx}
\usepackage{amsmath}

\begin{document}

\title{Optimal Stabilization of Boolean Networks through Collective Influence}%
\author{Jiannan Wang }
\email{wangjiannan@buaa.edu.cn}
\affiliation {School of Mathematics and Systems Science, Beihang University, Beijing, China}
\affiliation{Key Laboratory of Mathematics Informatics Behavioral Semantics, Ministry of Education, China}
\author{Sen Pei}
\affiliation{Department of Environmental Health Sciences, Mailman School of Public Health, Columbia University, New York, NY, USA}
\author{Wei Wei}
\affiliation {School of Mathematics and Systems Science, Beihang University, Beijing, China}
\affiliation{Key Laboratory of Mathematics Informatics Behavioral Semantics, Ministry of Education, China}
\author{Xiangnan Feng}
\affiliation {School of Mathematics and Systems Science, Beihang University, Beijing, China}
\affiliation{Key Laboratory of Mathematics Informatics Behavioral Semantics, Ministry of Education, China}
\author{Zhiming Zheng}
\affiliation {School of Mathematics and Systems Science, Beihang University, Beijing, China}
\affiliation{Key Laboratory of Mathematics Informatics Behavioral Semantics, Ministry of Education, China}
\begin{abstract}
The stability of Boolean networks has attracted much attention due to its wide applications in describing the dynamics of biological systems. During the past decades, much effort has been invested in unveiling how network structure and update rules will affect the stability of Boolean networks. In this paper, we aim to identify and control a minimal set of influential nodes that is capable of stabilizing an unstable Boolean network. By minimizing the largest eigenvalue of a modified non-backtracking matrix, we propose a method using the collective influence theory to identify the influential nodes in Boolean networks with high computational efficiency. We test the performance of collective influence on four different networks. Results show that the collective influence algorithm can stabilize each network with a smaller set of nodes than other heuristic algorithms. Our work provides a new insight into the mechanism that determines the stability of Boolean networks, which may find applications in identifying the virulence genes that lead to serious disease.
\end{abstract}
\maketitle

\section{Introduction}

Boolean network model was proposed by Kauffman in 1969 \cite{origin}. Unlike other models such as differential equations, Boolean network models the dynamics of a gene as a binary (on or off) switch, while the interactions among the genes are represented by a series Boolean functions. The large number of parameters concerning the details of the interactions are neglected in Boolean network model, which is a great simplification. Despite the simplification, Boolean network model is still able to provide a deep insight into the dynamics of biochemical systems \cite{cell1,cell2,regulatory1,regulatory2,regulatory3,application1,application2}, social networks \cite{spread1,spread2} and economic systems \cite{application3,application4}. Due to their wide applications, Boolean network models with various topology and update functions have attracted the attention of researchers during the past decades \cite{research1,research2,research3,research4}.

As the state space of a Boolean network is finite, there must be a time when the system gets back to one of its previous states. Given that the update functions are fixed, the Boolean network will finally evolve along a cycled orbit, which is called an attractor. The attractor is vital in the study of Boolean networks because of its close relationship with numerous phenomena in real complex systems. For example, in intracellular regulation dynamics, attractors are interpreted as different states of cell life cycle, such as growing, resting, division and dying; In multicellular organisms, such as humans, attractors correspond to the varieties of cells generated from cell differentiation \cite{origin}. One of the key problems about the attractors is the stability, which is the ability that they eliminate small perturbations as the system evolves. In an unstable network, the whole system can be influenced by a very small perturbation. In this situation, the dynamics of the network is affected dramatically, usually leading to a system failure. The stability of Boolean networks is crucial to understand the regulation of gene networks, as previous research indicates that real genetic regulatory networks usually lie on the critical region between stable and unstable \cite{origin}, so that creatures can not only survive most genetic mutations but maintain their diversity as well. Besides, the problem of stability is also relevant to certain kinds of cancer \cite{semiannealed}, since cells in cancer tissue exhibit much higher heterogeneity than normal \cite{cancer}.

During the past decades, researchers have made great progress in studying the stability of Boolean networks. In 1985, Derrida \textit{et al.} proposed the annealed approximation to predict the stability of their random Boolean network model \cite{annealed}. Since then, other researchers have applied the annealed approximation to networks with various degree distributions and update functions \cite{topology1,topology2,update1,update2}. However, there are far more topological features that could not be described solely by degree distributions, such as community structure, degree assortativity, and reciprocity. It was not until 2009 when Pomerance \textit{et al.} proposed the semi-annealed approximation that we are finally able to predict the stability of Boolean networks with arbitrary topology \cite{semiannealed}. The semi-annealed approximation greatly broadens the range of the research of Boolean networks. With this method, researchers are able to study the stability of a variety of Boolean network models constructed from real biological systems. For example, Moreno \textit{et al.} studied the stability of Boolean multilevel networks \cite{multiplex} and Edward Ott \textit{et al.} explored the joint effects of topology and update rules on the stability of Boolean networks \cite{joint}.

Previous research articles mainly study the stability of Boolean networks macroscopically, aiming to find the hidden criterion that determines the stability. However, due to the vast heterogeneity in topological and dynamical properties among the nodes in real complex networks, it is likely that the influence of each node on the network is quite different and a small fraction of nodes can make a disproportionate effect on the stability of the whole system. Identifying the influential nodes could contribute to a series of problems, such as the detection of virulence genes and the optimal immunization of epidemic spread \cite{vital1,vital2,vital3,vital4,vital5,vital6,vital7}. Previous articles illustrate that perturbation on frozen nodes and nodes in self-freezing loops can only affect the dynamics of Boolean networks temporarily, while the set of relevant nodes is the core that finally decides the attractor \cite{addition1,addition2,addition3,addition4}. In this paper, our purpose is to identify the minimum set of nodes through the control of which an unstable network can be stabilized. First we construct a theoretical framework to give a mathematical description of this problem. Then we transform this problem to the minimization of the largest eigenvalue of a modified non-backtracking matrix. Following the collective influence theory designed for optimal percolation \cite{vital1,add1,add2,vital3}, we propose the method of collective influence to identify the minimum set of influential nodes. Different from other methods such as PageRank \cite{pagerank} and k-core \cite{kcore}, the proposed collective influence takes into consideration not only the topological properties, but also the dynamical properties of nodes with respect to stability. Simulation results on four different Boolean network models show that our method outperforms the traditional benchmark methods, identifying a smaller set of nodes which is capable of achieving stabilization. What's more, our method also has superior computational efficiency.

\section{Stability of Boolean Networks}

A Boolean network consists of $N$ nodes connected with $L$ directed edges. For each node $i$, its state, often denoted as $x_i$, can only be one of the following two states: \textit{on} (1) or \textit{off} (0). The interactions among the \textit{N} nodes are defined by $N$ Boolean functions $\{ f_1,f_2,...f_N\}$, which are often referred to as the update functions. At time \textit{t}, the state of the network can be represented as an $N$-dimensional vector ${X(t)}=(x_1(t),x_2(t),...x_N(t))$. At time $t+1$, its state $X(t+1)$ is determined by $X(t)$ and the update functions. Take node $i$ for an example. Given that it has ${k_i}$ inputs ${\{i_1,i_2...i_{k_i}\}}$, its state at time $t+1$ is then given by the update function $f_i$ as:
\begin{equation}
x_i(t+1)=f_i(x_{i_1}(t),x_{i_2}(t),...x_{i_{k_i}}(t)).
\end{equation}
If we denote the set of nodes that input into $i$ as $ X_i=(x_{i_1},x_{i_2},...x_{i_{k_i}})$, the dynamics of the Boolean network can be represented as
\begin{equation}
x_i(t+1)=f_i(X_i(t)).
\end{equation}
The topology of the Boolean network is represented by the adjacent matrix $A$, whose elements $A_{ij}=1$ if there exists an edge that points from node $j$ to $i$, otherwise $A_{ij}=0$. To define the stability of Boolean networks, we consider two initial states of the network $X(t_0)$ and $\tilde{X}(t_0)$. The Hamming distance between these two states is defined as
\begin{equation}
H(t_0)=\frac{1}{N}\sum_{i=1}^N{|x_i(t_0)-\tilde{x}_i(t_0)|}.
\end{equation}

Let us suppose that $N$ is large enough and the initial Hamming distance between $X(t_0)$ and $\tilde{X}(t_0)$ is close, which means $H(X(t_0),\tilde{X}(t_0))\ll1$. The stability of Boolean networks mainly concerns the behavior of $H(t)$ as $t\rightarrow\infty$. The network is stable if $\lim_{t\rightarrow\infty}H(t)=0$, which indicates the system will eventually go back to normal despite small perturbations. Otherwise if $\lim_{t\rightarrow\infty}H(t)>0$, we regard the network as unstable. An unstable network usually fails to recover from a small perturbation spontaneously, hence it is easy for an external input to affect its dynamics dramatically.

\section{Optimal Stabilization through Collective Influence}

In this section, we discuss the optimal stabilization problem of Boolean networks. Previous research articles studying the stability of Boolean networks mainly focus on the criterion between stable networks and unstable ones \cite{topology1,update1,multiplex}. In a stable Boolean network, small perturbations tend to vanish spontaneously with time evolves, while in an unstable network, it can cause remarkable influence on the dynamics of the whole system. Concerning the property of unstable Boolean networks, it attracts our attention that whether we could stabilize an unstable Boolean network by making some nodes immune to any possible perturbations. If we are able to grant immunity to all nodes in the network, it is quite obvious that every Boolean network could be stabilized. But what if we could only immunize a fraction of nodes? How do we stabilize a Boolean network with as small fraction of immunized nodes as possible? Considering the vast heterogeneity in topological and dynamical properties among nodes in real biological systems, we certainly could not expect the best performance by selecting the targets randomly. In this paper, our purpose is to identify the minimum set of nodes that is able to stabilize an unstable Boolean network. In the following work, we call this problem the optimal stabilization of Boolean networks, and the selected targets that are immune to perturbations are referred to as controllers.

We start by proposing a mathematical description of the optimal stabilization problem. We use $\mu_i$ to represent whether a node $i$ is a controller: $\mu_i=0$ if node $i$ is controlled, otherwise $\mu_i=1$. Therefore the vector $\mu=(\mu_1,\mu_2,...\mu_N)$ contains the information of the selected controllers and we call it a configuration of the network. The fraction of controllers in the network is represented by
\begin{equation}
q=1-\frac{1}{N}\sum^{N}_{i=1}{\mu_i}=1-\langle \mu \rangle.
\end{equation}
Regarding the stability of Boolean networks, we represent the Hamming distance $H$ as a function of $q$:
\begin{equation}
H(q)=\lim_{t\rightarrow\infty}H(q,t)=\lim_{t\rightarrow\infty}\frac{1}{N}\sum^{N}_{i=1}{(x_i(t)-\tilde{x_i}(t))}.
\end{equation}
In a stable Boolean network, we could always expect that $\langle H(q)\rangle=0$, where $\langle \cdot\rangle$ stands for the average over all initial values. In an unstable network, however, it always turns out that $\langle H(q,t)\rangle >0$. The optimal stabilization problem is to find the minimum fraction $q_c$ of nodes and the corresponding optimal configuration such that $\langle H(q_c)\rangle=0$:
\begin{equation}
q_c=min\{q\in[0,1]:\langle H(q) \rangle=0\}.
\end{equation}
For $q\geq q_c$, there exist a variety of configurations which are able to stabilize the Boolean network. In contrast, for $q < q_c$, the configuration that could stabilize the whole system does not exist. With $q$ decreases from $1$ to $0$, the number of configurations which satisfy $\langle H(q) \rangle=0$ also decreases and vanishes eventually at $q_c$.

Considering a single node in the network, it could be perturbed only when it is not a controller and at least one of its inputs has been perturbed. Therefore, the parameter $\mu_i$ itself fails to measure the influence of perturbation. So we need another variable encoding the information that whether a node is perturbed or not. This information is stored in the variable $\nu_i$: $\nu_i = 1$ if node $i$ is perturbed, otherwise $\nu_i = 0$. The influence of the perturbation in the whole system is then represented by the fraction of perturbed nodes, when $q$ fraction of the nodes in the configuration $\mu$ are controlled:
\begin{equation}
H(q,\mu)=\frac{1}{N}\sum_{i=1}^N\nu_i.
\end{equation}
For a given fraction of controllers, the optimal stabilization of Boolean networks requires to minimize the influence of perturbation over all of the configurations. However, an explicit function of $H(q,\mu)$ is not available. Thus it is difficult for us to select the important nodes in Boolean network by minimizing $H(q,\mu)$ directly. Our main idea is to transform the problem of optimal stabilization into minimizing the largest eigenvalue of a modified non-backtracking matrix, which can be represented analytically.

To derive the relation between $\nu=(\nu_1,\nu_2,... ,\nu_N)$ and $\mu= (\mu_1,\mu_2,...\mu_N)$, we consider a directed edge that points from node $i$ to $j$. Let us suppose that node $j$ is temporarily removed from the network and we concern whether node $i$ is perturbed or not. This information is stored in variable $\nu_{i\rightarrow j}$, which represents the probability that node $i$ is perturbed in the absence of $j$. Clearly we can conclude that $\nu_{i\rightarrow j}=0$ if $\mu_i = 0$. So we only consider the case when $\mu_i = 1$. Given that $j$ is temporarily removed from the network, node $i$ is perturbed only because of the event "at least one of the nodes that point to node $i$ other than $j$ is perturbed". Given that topology of the network is locally tree-like, the variables $\nu_{i\rightarrow j}$ satisfy a closed set of equations:
\begin{equation}\label{main}
\nu_{i\rightarrow j}=\rho_i\mu_i(1-\prod_{k\in {\partial i \setminus j}}(1-\nu_{k\rightarrow i})),
\end{equation}
where $\rho_i$ represents the sensitivity of node $i$ and $\partial i$ is the set of nodes that input into $i$. Obviously, the system defined in Equation \ref{main} admits the solution $\{\nu_{i\rightarrow j}=0\}$ for all $i,j$. As a result, the impact of the perturbation in the whole network $H(q,\mu)=0$. By linearizing Equation \ref{main} and neglecting the terms whose orders are higher than one, we conclude that the stability of the solution $\{\nu_{i\rightarrow j}=0\}$ depends on the largest eigenvalue of the linear operator of the $L\times L$ matrix:
\begin{equation}
\hat{M}_{k\rightarrow l,i\rightarrow j}=\frac{\partial \nu_{i\rightarrow j}}{\partial \nu_{k\rightarrow l}}.
\end{equation}

We use $\lambda(q,\mu)$ to represent the largest eigenvalue of $\hat{M}$, which depends on the fraction of controllers $q$ and the configuration of the network $\mu$. According to the Frobenius theorem \cite{theorem}, the largest eigenvalue of $\hat{M}$ is real and positive. The stability of the solution $H(q,\mu) = 0$ is determined by the critical condition $\lambda(q_c,\mu^*)=1$, where $\mu^*$ is the optimal configuration. When $q < q_c$, for each configuration $\mu$ it turns out $\lambda(q,\mu)> 1$, hence it is impossible to find a set of controllers such that $H(q,\mu)=0$. On the contrary, when $q > q_c$, there are two different possibilities. On the one hand, for some non-optimal configurations we have $\lambda(q,\mu)>1$, which are unable to stabilize the whole network; On the other hand, there exist configurations that satisfy $\lambda(q,\mu)<1$, which corresponds to a stable solution of $H(q,\mu) =0$. As we approach from above, $q\mapsto q_c^+$, the number of configurations such that $\lambda(q,\mu)<1$ gradually decreases and eventually vanishes at $q_c$. The elements of the linear operator $\hat{M}$ can be represented in terms of the non-backtracking matrix $\hat{B}$ \cite{nonback,trackmatrix}:
\begin{equation}
\hat{M}_{k\rightarrow l,i\rightarrow j}=n_i\rho_i\hat{B}_{k\rightarrow l,i\rightarrow j},
\end{equation}
where
\begin{equation}\hat{B}_{k\rightarrow l,i\rightarrow j}=
\begin{cases}
1 & \text{if}\  l=i \ \text{and}\  j\neq k \\
0 & \text{otherwise}.
\end{cases}
\end{equation}

Due to the complexity of the dynamics of Boolean networks, it is difficult to give an analytical form of $\lambda(q,\mu)$. Our approach is to approximate $\lambda(q,\mu)$ with the Power Method, which is supposed to converge to its exact solution after sufficient steps of iterations. For a fixed configuration, we could use $\lambda(\mu)$ to represent the largest eigenvalue of $\hat{M}$ and the parameter $q$ can be omitted. Let us consider an arbitrary non-zero vector $\omega_0$. For convenience, we suppose that $\omega_0=(1,1,....1)^{t}$ and we use $\omega_l$ to represent the result of $\omega_0$ after $l$ iterations by $\hat{M}$:
\begin{equation}
\omega_l=\hat{M}^l\omega_0.
\end{equation}
According to the Power Method, the largest eigenvalue of $\hat{M}$ decides the growth rate of $\omega_0$, thus we can approximate the value of $\lambda(\mu)$ by calculating the growth rate of $\omega_l$ as $l\rightarrow \infty$:
\begin{equation}
\lambda(\mu)=\lim_{l\rightarrow \infty}\lambda_l(\mu)=\lim_{l\rightarrow \infty} (\frac{|w_l(\mu)|}{|w_0|})^{\frac{1}{l}}
\label{one}
\end{equation}

Our next step is to approximate the value of $\lambda_l(\mu)$, which eventually converges to the largest eigenvalue of $\hat{M}$. The elements of $\hat{M}$ contain the information of the connections of edges in Boolean networks, and its indices $k\rightarrow l$ and $i\rightarrow j$ correspond to directed edges. For computation convenience, we can embed it into an $N \times N\times N\times N$ matrix $M$:
\begin{equation}
M_{ijkl}=\mu_k\rho_kA_{ij}A_{kl}\delta_{jk}(1-\delta_{il}),
\end{equation}
where $i,j,k$ and $l$ all vary from $1$ to $N$. As for the $L$ dimensional initial vector $\omega_0$ whose elements are all $1$, its projection in the enlarged $N \times N\times N\times N$ space is an $N \times N$ vector, whose elements are $|\omega_0\rangle=A_{ij}$. Given the topology of the Boolean network, we can calculate the right vector $|\omega_1(\mu)\rangle$ as
\begin{equation}
|w_1\rangle_{ij}=\sum_{kl}M_{ijkl}|w_0\rangle_{kl}=\mu_j\rho_jA_{ij}K_j^{out},
\end{equation}
while the left vector $\langle w_1(n)|$ can be calculated as
\begin{equation}
_{ij}\langle w_1|=\sum_{kl} {_{kl}\langle w_0|M_{klij}=\mu_i\rho_iA_{ij}K_i^{in}}.
\end{equation}
With the left and right vectors above, we can write the expression of the norm of $|w_1(\mu)|$ as
\begin{equation}
|w_1(\mu)|^2=\sum_{ij}{_{ij}\langle w_1|w_1\rangle_{ij}}=\sum_{ij}\mu_i\mu_j\rho_i\rho_jA_{ij}K_i^{in}K_j^{out}.
\end{equation}
The norm of $\omega_0$ can be easily calculated as
\begin{equation}
|w_0(\mu)|^2=\sum_{ij}{_{ij}\langle w_0|w_0\rangle_{ij}}=L.
\end{equation}
According to Equation \ref{one}, we can finally give the mathematical formula of $\lambda_1(\mu)$:
\begin{equation}
\lambda_1(\mu)=(\frac{1}{L}\sum_{ij}\mu_i\mu_j\rho_i\rho_jA_{ij}K_i^{in}K_j^{out})^{\frac{1}{2}}.
\end{equation}
Similarly, we can calculate the right vector $|w_2(\mu)\rangle$ as
\begin{equation}
|w_2\rangle_{ij}=\sum_{kl}M_{ijkl}|w_1\rangle_{kl}=\mu_j\rho_jA_{ij}\sum_lA_{jl}\mu_l\rho_lK_l^{out}(1-\delta_{il}),
\end{equation}
while its left vector $\langle w_2(\mu)|$ is given by
\begin{equation}
_{ij}\langle w_2|=\sum_{kl} {_{kl}\langle w_1|M_{klij}=\mu_i\rho_iA_{ij}\sum_k\mu_k\rho_kA_{ki}K_k^{in}(1-\delta_{jk})}.
\end{equation}
Thus we can calculate the norm of $|w_2(\mu)|$:
\begin{equation}
\begin{split}
|w_2(\mu)|^2=&\sum_{ij}{_{ij}\langle w_2|w_2\rangle_{ij}}\\
=&\sum_{ijkl}\mu_i\rho_i\mu_j\rho_j\mu_k\rho_k\mu_l\rho_lA_{ij}A_{jk}A_{kl}K_i^{in}K_l^{out}\\
&(1-\delta_{ik})(1-\delta_{jl}).
\end{split}
\end{equation}
We are able to represent $\lambda_2(\mu)$ as
\begin{equation}
\begin{split}
\lambda_2(\mu)=&(\frac{1}{L}\sum_{ijkl}\mu_i\rho_i\mu_j\rho_j\mu_k\rho_k\mu_l\rho_lA_{ij}A_{jk}A_{kl}K_i^{in}K_l^{out}\\
&(1-\delta_{ik})(1-\delta_{jl}))^{\frac{1}{4}}.
\end{split}
\end{equation}

From the formulas of $\lambda_1(\mu)$ and $\lambda_2(\mu)$, we can write down higher orders of iterations $\lambda_n(\mu)$ \cite{vital1}. The order of interaction is defined as the number of node appearing in it. In the case $l=1$, $\lambda_1$ corresponds to a two-body iteration problem concerning $i$ and $j$. In the iteration, since node $i$ points a directed edge to node $j$, the variables $\mu_i\rho_i$ and $\mu_j\rho_j$ are multiplied by each other. This factor is then multiplied by the in-degree of the initial node and the out-degree of the ending node of the edge, which is just equal to $\mu_i\rho_i\mu_j\rho_jK_i^{in}K_j^{out}$. As for the case $l=2$, we can find similarly that $\lambda_2$ corresponds to a $l=2$ non-backtracking walk on Boolean networks, which is a four-body interaction problem if $i\neq l$, or a three-body interaction problem when $i=l$. Here, we can see that the series expansion of the maximum eigenvalue can be written in terms of a systematic diagrammatic expansion of increasing levels of multi-body interactions. Typically, $\lambda_l$ corresponds to a non-backtracking walk of length is $l$, which could involve as much as $2l$ nodes in the interaction. We have to stress that the initial and final nodes of the non-backtracking walks do not necessarily need to be different, since loops are allowed in non-backtracking walks. However, due to the fact that most networks in real world are sparse and locally tree like, we decide that those non-backtracking walks with loops are negligible and each non-backtracking walk from node $i$ to $j$ is in fact a shortest path between the two nodes. Thus the norm of $|w_l(\mu)|$ can be represented as:
\begin{equation}
|w_l(\mu)|^2=\sum_{i=1}^N K_i^{in}\sum_{j\in\partial Ball(i,l)}(\prod_{k \in P_l(i,j)}\mu_k\rho_k) K_j^{out},
\end{equation}
where $Ball(i,l)$ consists of the nodes within a ball of radius $l$ from node $i$(defined as the shortest path), $\partial Ball(i,l)$ is the surface of the ball and $P_l(i,j)$ is the shortest directed path of length $l$ from node $i$ to $j$. Here, we define the collective influence of node $i$ as
\begin{equation}
CI_l(i)=K_i^{in}\sum_{j\in\partial Ball(i,l)} (\prod_{k \in P_l(i,j)}\mu_k\rho_k)K_j^{out}.
\end{equation}
Therefore, the norm $|w_l(n)|^2$ is the sum of collective influence of all nodes:
\begin{equation}
|w_l(\mu)|^2=\sum_{i=1}^N CI_l(i).
\end{equation}

To minimize $\lambda(\mu)$, our main idea is to use the greedy algorithm and select one controller at a time. At each step, we select the node with the highest score of collective influence as a controller, which results in the biggest drop in the value of $|w_l(n)|^2$. The selected controller is virtually removed from the network and the topology of the network is updated. Then we continue to select the next controller until the network is finally stabilized. One advantage of using the greedy algorithm is that every time when a controller is selected, the factors $\mu_i$ in the formula of collective influence can be just ignored. Since the selected controllers are virtually removed, $\mu_i=1$ for every remaining node in the network. Another problem of collective influence is that a proper radius $l$ still needs to be defined. Intuitively the performance of collective influence is better with a larger radius $l$, but in the mean time the complexity of computing the score of collective influence increases dramatically. When we consider the case $l=0$, we get $CI_0(i)=K_i^{in}q_i$, which is the same as the high degree strategy. Previous articles have proved this strategy to be less than satisfied \cite{vital5}. Thus we go further to consider the case of $l=1$, where the collective influence is given by
\begin{equation}
CI_1(i)=K_i^{in}\rho_i\sum_{j\in\Gamma i}\rho_jK_j^{out},
\end{equation}
where $\Gamma_i$ is the set of nodes that node $i$ points to. The score of $CI_1(i)$ consists of not only the in-degree of node $i$ itself, but also the topological information of its nearest neighbours. We can expect better performances when considering larger radii, but the computation will be more time consuming.

In general, the collective influence algorithm is scalable for large networks with a computational complexity $O(N\log N)$. Computing the collective influence is equivalent to iteratively visiting subcritical neighbors of each node layer by layer within a radius of $l$. Since $l$ is finite, it takes $O(1)$ time to compute the collective influence of each node. Initially, we have to calculate the collective influence for all nodes in the network. However, during later steps, we only have to recalculate for nodes within a $l+1$ radius from the selected controllers, which scales as $O(N)$. When it comes to selecting the node with the highest collective influence, we can make use of the data structure of heap that takes $O(\log N)$ time. Therefore, the overall complexity of the collective influence algorithm is $O(N\log N)$. In the following section, we mainly take $CI_1$ as the representative of collective influence and discuss its performance compared by other algorithms.

\section{Numerical Simulations}

In this section, we construct four different Boolean networks on which we can test the performance of collective influence. First we consider Kauffman's original $N-K$ network model, where all nodes have exactly $K$ inputs randomly selected from the other $N-1$ nodes. The degree distribution of $N-K$ network model does not show much heterogeneity, which is quite different from real genetic regulatory networks. Considering that real genetic regulatory networks usually have short tailed in-degree distributions and long-tailed out-degree distributions \cite{distribution1,distribution2}, we construct the second network with Poisson distributed in-degrees and scale-free out-degrees. These two networks above are typical toy models of Boolean network. The following two networks are constructed using data from real world systems. The third network captures the innovation spread among $241$ physicians in four towns of Illinois, Peoria, Bloomington, Quincy and Galesburg \cite{source1}. The last network was created from a survey on the social relationship among adolescents \cite{source2}. On each of the four networks above, the update functions are given in the form of a truth table. For any inputs of $f_i$, its corresponding output is randomly chosen from $\{0,1\}$ with probability $0.5$. As a result, all nodes in the networks have a common sensitivity $\rho_i\equiv \rho \equiv0.5$.

On each network, we compare the performance of collective influence with the following methods: high degree (HD) \cite{highdegree}, eigenvector centrality (EC) \cite{eigenvector}, Google PageRank (PR) \cite{pagerank} and voter rank (VR) \cite{vital4}, which have been proved useful in detecting influential nodes in complex networks. High degree strategy, as its name implies, defines the influence of a node $i$ with its degree. Here, we choose $K_i^{in}$ as the degree rank of node $i$. To get better performances, here we adopt the adaptive version of high degree method (HDA). After each selection, the degree of each node is recalculated. However, as it has been mentioned in various studies, nodes with high degree don't necessarily possess high influence. The eigenvector centrality fixes this problem by considering not only the number of a node's neighbours, but also the influence of its neighbours, known as the mutual enhancement effect. For each node $i$, its score of eigenvector centrality $EC_i$ is given by
\begin{equation}
EC_i=\sum_j A_{ij} EC_j.
\end{equation}
PageRank is a famous algorithm that is used to rank websites in google search engines and other commercial scenarios. According to the Page rank algorithm, the influence of a webpage is determined by random walking on the network constructed from the relationships of web pages. Mathematically, the PageRank score of node $i$ is
\begin{equation}
PR_i=\sum_j A_{ij} \frac{PR_j}{k_j^{out}}.
\end{equation}
Due to the wide existence of community structure in complex networks, influential nodes in complex networks are more likely to connect with each other, which result in the fact that their sphere of influence tend to overlap. To avoid such conditions, in the voter rank algorithm, each node is granted an initial voting ability $\theta_i$ and the score of voter rank is calculated as:
\begin{equation}
VR_i=\sum_j A_{ij} \theta_j.
\end{equation}
At each step, the voter rank select one single node with the highest voting score. Then the vote abilities of its neighbours spontaneously decrease. Therefore the nodes nearby are less likely to be chosen in the following process and the selected nodes are less likely to be close to each other. For each algorithm above, we start from $q=0$ and pick the controllers one after another until $q=0.2$. Every time a controller is chosen, we calculate the average Hamming distance $\langle H \rangle$ to see whether or not the network has been stabilized.

\begin{figure}[htbp]
  \centering
  \includegraphics[width=1\columnwidth]{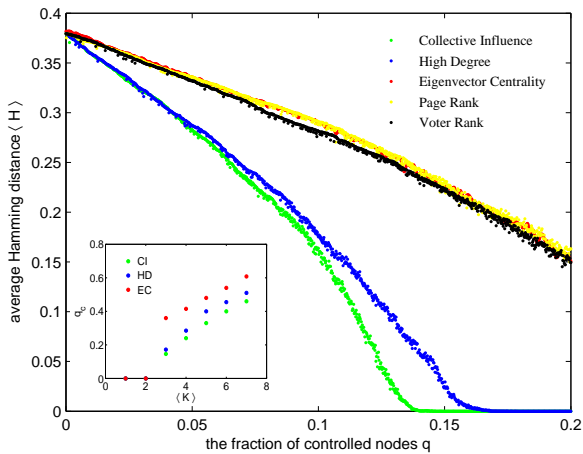}
  \caption{Normalized average Hamming distance $\langle H \rangle$ plotted against the fraction of controllers $q$ in $N-K$ network. The network consists of 1000 nodes and its average degree equals to 3. The performances of collective influence, high degree, eigenvector centrality, PageRank and voter rank are respectively represented in green, blue, red, yellow and black. The small panel shows the results of $q_c$ with the increase of the average degree K.}
  \label{1}
\end{figure}

\begin{figure}[htbp]
  \centering
  \includegraphics[width=1\columnwidth]{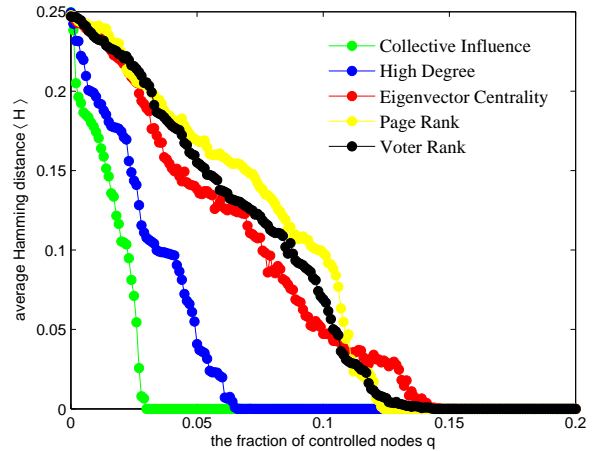}
  \caption{Normalized average Hamming distance $\langle H \rangle$ plotted against the fraction of controllers $q$ in a heterogeneous network. The networks is constructed according to the configuration model. The in-degrees of nodes follow Poisson distribution and the out-degrees are scale-free. The network consists of 1000 nodes and its average degree equals to 3. The performances of collective influence, high degree,eigenvector centrality, PageRank and voter rank are respectively represented in green, blue, red, yellow and black.}
  \label{2}
\end{figure}

As for the calculation of the Hamming distance $\langle H \rangle$, we take it as the average of 100 pairs of initial values. For each initial value, $H$ is calculated through the following procedure. First, we randomly generate an initial value $X(t)$ and evolve it according to update functions till $t_0=100$, where we expect it to have completed any transient behaviors. Next, we chose a small fraction $(\varepsilon=0.01)$ of its components and flip their states to create a perturbed value $\tilde{X}(t_0)$. In other words, $\tilde{x_i}(t_0)=1-x_i(t_0)$ if node $i$ is perturbed, otherwise $\tilde{x_i}(t_0)=x_i(t_0)$. The initial Hamming distance is $H(X(t_0),\tilde{X}(t_0))=0.01$. Finally, we take $X(t_0)$ and $\tilde{X}(t_0)$ as the initial values and evolve both of them in parallel. Here we stress that for nodes that has been chosen as controllers, their states in both orbits are always the same, since controllers are immune to any perturbations. Our main interest lies on the long-time behavior of $H$, which is calculated by averaging $H(X(t),\tilde{X(t)})$ from \textit{t=400} to \textit{t=500}. This whole procedure is repeated for 100 times and $\langle H \rangle$ is the average of $H$ over all initial values.

From Fig.\ref{1} we can see the performance of the five algorithms mentioned above in Kauffman's $N-K$ network model. The network consists of $N=5000$ nodes and its average degree $K=3$. For each algorithm, the average Hamming distance $\langle H \rangle$ decreases with the increase of the fraction of controllers $q$. During the process, collective influence outperforms the other four algorithms and stabilizes the network with the minimal fraction of controllers $q_c^{CI}\thickapprox 0.13$, followed by high degree strategy that stabilizes the network at $q_c^{HD}\thickapprox 0.17$. As for the other three algorithms, their performances are rather close to each other. The average Hamming distance remains $\langle H \rangle \thickapprox 0.15$ even when $q=0.2$ of nodes in the network are under control. One interesting phenomenon shown in Fig.\ref{1} is that the performance of high degree beats those more complex algorithms like eigenvector centrality, PageRank and voter rank, which usually perform quite well. One possible reason could lie in the difference between optimal stabilization of Boolean networks and those rank problems that these algorithms are designed for. According to these three algorithms, a node usually exhibits higher importance if it is pointed by more nodes with higher importance themselves. However, in the problem of optimal stabilization, it's pretty much the opposite. In this problem, one node enjoys higher influence by pointing to more nodes with higher influence. The small panel shows that with the average degree of the $N-K$ increases, the minimal fraction of controllers $q_c$ increases as well. During the process, the collective influence still outperforms the other algorithms.
\begin{figure}[htbp]
  \centering
  \includegraphics[width=1\columnwidth]{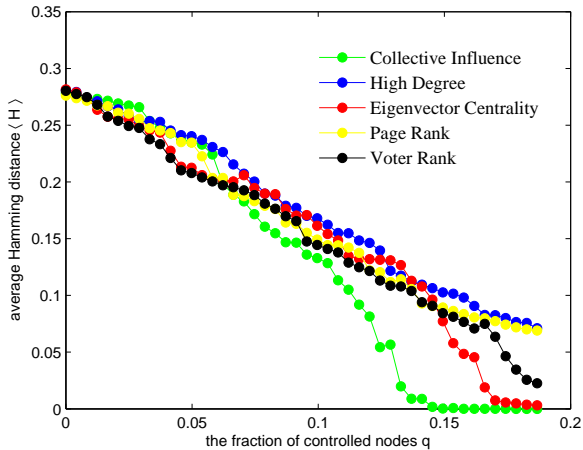}
  \caption{Normalized average Hamming distance $\langle H \rangle$ plotted against the fraction of controllers $q$ in the Physician network. The network consists of 241 nodes and 1098 edges. The performances of collective influence, high degree, eigenvector centrality, PageRank and voter rank are respectively represented in green, blue, red, yellow and black. Simulations are performed for an initial Hamming distance $H(t_0)=0.01$ and the results of $\langle H \rangle$ are averaged over 100 random initial values.}
  \label{3}
\end{figure}

\begin{figure}[htbp]
  \centering
  \includegraphics[width=1\columnwidth]{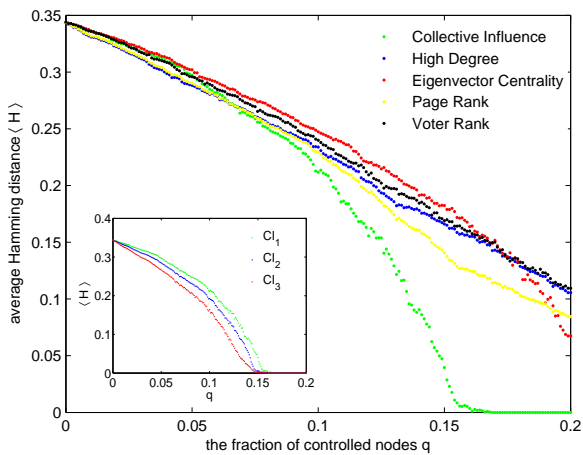}
  \caption{Normalized average Hamming distance $\langle H \rangle$ plotted against the fraction of controllers $q$ in student social network. The network consists of 1000 nodes and 4175 edges, which is part of the adolescent social network constructed according to the survey. The performances of collective influence, high degree,eigenvector centrality, PageRank and voter rank are respectively represented in green, blue, red, yellow and black. The small panel shows the performance of the collective influence algorithm when larger radii are applied.}
  \label{4}
\end{figure}

In Fig.\ref{2}, we show the performances of the five algorithms on a heterogeneous network. The networks is constructed using the configuration model, which consists of $1000$ nodes and its average degree is $3$. From Fig.\ref{2} we can see that with the increase of the fraction of controllers $q$, the network reaches stable region first at $q_c^{CI}\thickapprox 0.03$ when collective influence is applied. Again, the high degree strategy outperforms the other three algorithms and stabilizes the network at $q_c^{HD}\thickapprox 0.07$. The performances of eigenvector centrality, PageRank and voter rank are pretty similar, which could not stabilize the network until $q=0.14$. When we compare Fig.\ref{1} and Fig.\ref{2}, it is interesting that although these two networks have the same average degree, an algorithm can behave quite differently on the two networks. In general, an algorithm can achieve stabilization with a much smaller fraction of controllers in the heterogeneous network, since it is easier to control the dynamics of the whole system if the network exhibits more topological heterogeneity. In Kauffman's $N-K$ network model, however, the importance of nodes is quite similar to each other, thus it is unlikely to achieve stabilization by simply controlling a small fraction of them.

Fig.\ref{3} and Fig.\ref{4} show the performances of the five algorithms when applied into real-world networks. In Fig.\ref{3}, each node in the network represents a physician and each directed edge from node $i$ to node $j$ shows that physician $i$ regards physician $j$ as his friend or he turns to $j$ if he needs advice or is interested in a discussion. There always only exists one edge between two nodes even if more than one of the listed conditions are true. The network in Fig.\ref{4} is created from a survey including $2539$ students. In the survey, each student was asked to list his five best female and five male friends. Each node represents a student and an edge from node $i$ to $j$ means that student $i$ chose student $j$ as a friend. In the simulation, we chose part of the network that contains $1000$ students among them. From Fig.\ref{3} and Fig.\ref{4}, we can see that collective influence outperforms the other four algorithms in both networks. In Fig.\ref{3}, the Physician network becomes stable at $q_c^{CI}\thickapprox 0.15$ when collective influence is applied, followed by eigenvector centrality which achieves stable region at $q_c^{EC}\thickapprox 0.18$. According to Fig.\ref{4}, collective influence is the only algorithm that is able to achieve stabilization of adolescent social network with $q_c^{CI}\thickapprox 0.17$, while for the other four algorithms, the network will not be stabilized even at $q=0.2$. We continue to compare the performances of $CI1$, $CI_2$ and $CI_3$, the results in the small show that the performance of collective influence algorithm improves with a larger radius $l$, but the improvement is rather limited.

\section{Conclusion}

In this paper, we propose the optimal stabilization problem of Boolean networks, which aims to identify the minimal set of influential nodes that is capable of stabilizing an unstable Boolean network. Since it is difficult to represent the average Hamming distance as a function of the configuration of the network, we transform this problem into minimizing the largest eigenvalue of a modified non-backtracking matrix that decides the stability of Boolean networks. We propose collective influence which enables us to identify the influential nodes with respect to the stability of Boolean networks. To test the performance of collective influence, we construct two toy networks and two real-world networks on which we compare the performance of collective influence with other four algorithms: high degree, eigenvector centrality, PageRank, and voter rank. The results show that in all four networks, our collective influence algorithm outperforms others by stabilizing the networks with a smaller fraction of controllers. We also find that it is easier to stabilize a Boolean network with more heterogeneity. Our work could contribute to identifying the virulence genes that cause serious inherited disease. Besides, it also provides a new insight into the mechanism that determines the stability of Boolean networks, which is useful to control the dynamics in a series of real biological systems.

\end{document}